\documentstyle[12pt]{article}

\input tcilatex
\QQQ{Language}{
American English
}

\begin{document}

\bigskip

\begin{center}
{\huge {\bf The Exact Evolution Equation}}

{\huge {\bf of the Curvature Perturbation}}

{\huge {\bf for Closed Universe}}

\bigskip \bigskip

{\Large {\bf De-Hai Zhang\footnote{~~dhzhang@gscas.ac.cn} and Cheng-Yi Sun}}%
\\[0pt]

\bigskip

Department of Physics,\\[0pt]

The Graduate School of The Chinese Academy of Sciences,\\[0pt]

P.O.Box 3908, Beijing 100039, P.R.China.\\[0pt]

\bigskip

{\bf Abstract:}
\end{center}

{As} is well known, the exact evolution equation of the curvature
perturbation plays a very important role in investigation of the inflation
power spectrum of the flat universe. However, its corresponding exact
extension for the non-flat universes has not yet been given out clearly. The
interest in the non-flat, specially closed, universes is being aroused
recently. The need of this extension is pressing. We start with most
elementary physical consideration and obtain finally this exact evolution
equation of the curvature perturbation for the non-flat universes, as well
as the evolutionary controlling parameter and the exact expression of the
variable mass in this equation. We approximately do a primitive and immature
analysis on the power spectrum of non-flat universes. This analysis shows
that this exact evolution equation of the curvature perturbation for the
non-flat universes is very complicated, and we need to do a lot of numerical
and analytic work for this new equation in future in order to judge whether
the universe is flat or closed by comparison between theories and
observations.

\bigskip



\bigskip

\section{ Introduction}

{\hspace*{5mm}}In recent years the cosmology acquires a dazzling
development. Unexpectedly, an elusory ingredient in the eyes of physicists,
i.e., a tiny non-zero vacuum energy or so-called ``dark energy'', turns out
to be a main component of our present universe. Some evidence for the once
recondite inflation has likely emerged now. A nice observed curve about the
power spectrum of the Cosmic Microwave Background Radiation (CMBR) is
actually fine consistent with the theoretical prediction\cite{WMAP}. The
best model may be the slow-roll inflation of the flat universe. All seem
faultless. A new era of the precision cosmology has been coming.

However, there is still some little bother. It seems the central value of
the total density of the universe is slightly larger, $\Omega _T=$$1.02\pm
0.02$. The large angle power spectrum looks like rather low\cite{Efsta}\cite
{COBE}. Whether or not these clues come into real questions? Maybe these are
only observed errors, the future experiments will tell us that the universe
is indeed flat, and the low power of the larger angle scales comes only from
imprecise data. However, no matter how precise our future data are, we shall
probably never be able to rule out the non-flat universe models. A couple of
years ago, when one didn't find out enough matter density in the universe
and didn't believe the cosmological constant will play important role in it,
the open universe models were once voguish. Recently, the possibility of the
closed universe model is being reconsidered\cite{Linde03}. Indeed, the idea
of the closed universe model has a long history \cite{Wheeler} \cite{Hawking}%
\cite{W-S}, some reasons seem attractive\cite{Linde95}. Anyhow, no matter
the universe is flat, open or closed, it is important for us that we must
know how to calculate exactly its primary power spectrum, in order for us to
do a rigour comparison between observations and theories.

As is well known, the inflation spectrum for the flat universe with single
inflaton field has been understood very well, which is expressed by an
evolution equation of the perturbation 
\begin{equation}
\label{(1)}\chi _k^{\prime \prime }+(k^2-z_0^{\prime \prime }/z_0)\chi
_k=0,\quad \quad z_0=a\dot \phi /H, 
\end{equation}
as well as the curvature spectrum 
\begin{equation}
\label{(2)}P_k^R=\frac{k^3}{2\pi ^2}\frac{|\chi _k|^2}{z_0^2}, 
\end{equation}
and forms a standard paradigm\cite{M-F-B}, where the variable mass $%
m_v^2=-z_0^{\prime \prime }/z_0$ has an exact expression even for a no
slow-roll case 
\begin{equation}
\label{(3)}z_0^{\prime \prime }/z_0=a^2H^2(2+2\epsilon _1-3\epsilon
_2+2\epsilon _1^2-4\epsilon _1\epsilon _2+\epsilon _3). 
\end{equation}
All notations can be found later. The fundamentality of all these results is
never overestimated in the analysis of the primary power spectrum.

However, for the non-flat case, especially for closed universe, the
corresponding exact equations have not been written down clearly up to now,
although some results are close to this goal nearly. Only when we know how
to extend these equations to the non-flat case, and make theoretical
predictions based on these new equations, we are able to judge whether the
universe was flat, open or closed by observational confirmations.

The goal of this letter is to provide the exact evolution equation of the
scalar perturbation for non-flat universes from some elementary physical
consideration. We do further some primitive, certainly immature,
approximative analysis based on our exact equations. This letter is
organized as follows. In section 2, we shall start with the perturbative
formalism of the Einstein equation to obtain the curvature perturbative
evolution equation for non-flat universes. In the section 3, we try to do an
analogic slow-roll analysis for these equations, look into whether the
slow-roll approximation is suitable. In the section 4, we want to do a brief
conclusion. Since our result may lay a foundation of further analysis for
non-flat universes in future, the correctness of our result is very key. So
it is necessary to give out the middle derivation as detailed as possible in
order for interested someone to check it conveniently. In this paper, we
only concentrate on derivation of the main formal results. A detailed
analysis of the perturbation spectrum in phenomenology will be the subject
of a forthcoming work.

\section{Perturbative evolution equation for non-flat universes}

{\hspace*{5mm}}Our starting point is the standard effective action of the
simple coupling system of Einstein gravity and single inflaton scalar field
with an arbitrary inflation potential $V(\phi )$, 
\begin{equation}
\label{1}S=\int \sqrt{g}[-\frac 1{16\pi G}R+\dfrac 12g^{\mu \nu }\partial
_\mu \phi \partial _\nu \phi -V(\phi )]d^4x. 
\end{equation}
We shall take $8\pi G=M_{pl}^{-2}=1$. Now, supposing that inflaton field
owns perturbation, we have 
\begin{equation}
\label{2}\phi (t,\vec x)=\phi _0(t)+\delta \phi (t,\vec x). 
\end{equation}
The perturbative metric in the longitudinal gauge, in terms of the Newtonian
gravitational potential $\Phi $, is 
\begin{equation}
\label{3}ds^2=[1+2\Phi (t,\vec x)]dt^2-[1-2\Phi (t,\vec x)]\frac{a^2(t)}{%
(1+K_cr^2/4)^2}(dr^2+r^2d\theta ^2+r^2\sin {}^2\theta d\varphi ^2). 
\end{equation}
The background motion equations are 
\begin{equation}
\label{4}H^2+\frac{K_c}{a^2}=\frac 13(\frac 12\dot \phi _0^2+V),\quad \quad
\ddot \phi _0+3H\dot \phi _0+V_{,\phi }=0,\quad \quad H\equiv \frac{\dot a}%
a, 
\end{equation}
hereafter a ``dot'' denotes derivative with respect to the physical time, $%
\dot f=df/dt$, and so on. The parameter $K_c$ takes $1$, $0$ or $-1$ for the
closed, flat or open universe respectively.

From the standard literatures, we can get the Einstein tensor and the
energy-momentum tensor in the perturbative form 
\begin{equation}
\label{5}\delta G_i^0=(\dot \Phi +H\Phi )_{,i},\quad \quad \delta
G_0^0=\dfrac 1{a^2}\Delta \Phi -3H\dot \Phi +3(\dfrac{K_c}{a^2}-H^2)\Phi , 
\end{equation}
\begin{equation}
\label{7}\delta T_i^0=\dot \phi _0^2(\dfrac{\delta \phi }{\dot \phi _0})_{,\
i},\quad \quad \delta T_0^0=\dot \phi _0^2[(\dfrac{\delta \phi }{\dot \phi _0%
})^{\bullet }-\Phi ]-3H\dot \phi _0^2(\dfrac{\delta \phi }{\dot \phi _0}). 
\end{equation}
Thus the linearized Einstein equation $\delta G_i^j=4\pi G\delta T_i^j$ for
the component of $0i$ is, 
\begin{equation}
\label{6}(a\Phi )^{\bullet }=\frac 12a\dot \phi _0^2(\dfrac{\delta \phi }{%
\dot \phi _0}). 
\end{equation}
Another Einstein equation for the component of $00$, having used (\ref{6}),
is 
\begin{equation}
\label{9}(\dfrac{\delta \phi }{\dot \phi _0})^{\bullet }=(1+\dfrac{2(\Delta
+3K_c)}{a^2\dot \phi _0^2})\Phi . 
\end{equation}
Both equations have been given in Ref.\cite{G-M}.

The motion equation of $\Phi $ is simple. In order to express this equation
in a more convenient way for later use we introduce a new variable $S\equiv
a\Phi $. Using (\ref{6}) and (\ref{9}) we have an equation with regard to $%
\Phi $ in fact 
\begin{equation}
\label{10}\ddot S=[\frac 12\dot \phi _0^2+a^{-2}(\Delta +3K_c)]S+\frac
d{dt}[\ln (a\dot \phi _0^2)]\cdot \dot S. 
\end{equation}
However, what we need really is the motion equation about the intrinsic
curvature perturbation $R$ of comoving hypersurfaces, which can expressed
as, seeing in Ref.\cite{Bardeen}, 
\begin{equation}
\label{11}R=\Phi +H\dfrac{\delta \phi }{\dot \phi _0}. 
\end{equation}
In order to gain the exact equation about $R$ in a clear way, we need to
define the rolling parameters, as we emphasize, which are not only valid for
slow-roll case, but also for no slow-roll case 
\begin{equation}
\label{12}\eta \equiv a^{-2}H^{-2},\quad \epsilon _1\ \equiv \ \dfrac{\dot
\phi _0^2}{2H^2},\quad \epsilon _2\equiv -\dfrac{\ddot \phi _0}{H\dot \phi _0%
},\quad \epsilon _3\equiv \dfrac{\phi _0^{(3)}}{H^2\dot \phi _0},\quad
\epsilon _4\equiv \dfrac{\phi _0^{(4)}}{H^3\dot \phi _0}. 
\end{equation}
The background motion equations (\ref{4}) can be rewritten as 
\begin{equation}
\label{13}H^2=\dfrac V3(1+K_c\eta -\frac 13\epsilon _1)^{-1},\quad \quad
\dot \phi _0=-\dfrac{V_{,\phi }}{3H}(1-\frac 13\epsilon _2)^{-1}, 
\end{equation}
which formalism is particularly suitable for parameter expansion if the
rolling parameters are small, in order to be used later. During the progress
of getting the equation about $R$, we shall meet many derivatives of these
various rolling parameters with respect to time, which should be listed
here, seeing Ref.\cite{S-G}, 
\begin{equation}
\label{14}\dot H=H^2(K_c\eta -\epsilon _1),\quad \quad \dot \eta =-2H\eta
(1+K_c\eta -\epsilon _1),\qquad \dot \epsilon _1=2H\epsilon _1(\epsilon
_1-\epsilon _2-K_c\eta ), 
\end{equation}
\begin{equation}
\label{15}\dot \epsilon _2=-H(\epsilon _3-\epsilon _1\epsilon _2-\epsilon
_2^2+K_c\epsilon _2\eta ),\qquad \dot \epsilon _3=H[\epsilon _4+\epsilon
_3(2\epsilon _1+\epsilon _2-2K_c\eta )]. 
\end{equation}
Then the Eq.(\ref{10}) can be rewritten as 
\begin{equation}
\label{17}\ddot S=H^2[\epsilon _1+\eta (-k^2+4K_c)]S+H(1-2\epsilon _2)\dot
S, 
\end{equation}
where we have use the eigenfunction expansion for some fluctuation $F(t,\vec
x)$ 
\begin{equation}
\label{18}F(t,\vec x)=\sum_{klm}F_{klm}(t)Q_{kl}(r)Y_{lm}(\theta ,\varphi ), 
\end{equation}
\begin{equation}
\label{19}(\Delta +k^2-K_c)Q_{kl}=0. 
\end{equation}
All definitions here can be found in Ref.\cite{A-S}. Strictly speaking, $%
\bar k\equiv \sqrt{k^2-K_c}$ is just the dimensionless comoving wavenumber
of a fluctuation. Hereafter our notations $S$ and $R$ mean their expanded
amplitudes $S_k(t)$ and $R_k(t)$, since the isotropy implies independent on
the indices $l$ and $m$.

Using Eq.(\ref{6}), the equation of curvature perturbation, i.e. Eq.(\ref{11}%
), is recast as 
\begin{equation}
\label{30}R=a^{-1}S+(aH\epsilon _1)^{-1}\dot S. 
\end{equation}
Doing derivative of it and using (\ref{17}), we get 
\begin{equation}
\label{31}\dot R=\eta (a\epsilon _1)^{-1}[-(k^2-4K_c)HS+K_c\dot S]. 
\end{equation}
Solve $R$ and $\dot R$ from above two equations, we have 
\begin{equation}
\label{32}S=a\epsilon _1(k^2-4K_c+K_c\epsilon _1)^{-1}[K_cR-aH\dot R], 
\end{equation}
\begin{equation}
\label{33}\dot S=aH\epsilon _1(k^2-4K_c+K_c\epsilon
_1)^{-1}[(k^2-4K_c)R+aH\epsilon _1\dot R]. 
\end{equation}
We can continue to apply derivative to (\ref{31}), again using (\ref{17}),
yield 
\begin{equation}
\label{34}\ddot R=(a^3\epsilon _1)^{-1}\{[K_c\epsilon
_1+(k^2-4K_c)(3+\epsilon _1-2\epsilon _2-2K_c\eta )]S-H^{-1}(k^2-2K_c)\dot
S\}. 
\end{equation}
In the following we want to use the conformal time $\tau =\int dt/a$, and
hereafter a ``prime'' denotes derivative with respect to the conformal time, 
$f^{\prime }=df/d\tau $, and so on. We have a transformation concerned $R$%
\begin{equation}
\label{35}\dot R=a^{-1}R^{\prime },\quad \quad \ddot R=a^{-2}R^{\prime
\prime }-a^{-1}HR^{\prime }. 
\end{equation}
Substituting Eqs.(\ref{32}) and (\ref{33}) into the right hand side of Eq.(%
\ref{34}), and changing to the conformal time, we establish 
\begin{equation}
\label{36}R^{\prime \prime }=-k^2R+C_0R+C_1R^{\prime }, 
\end{equation}
where the coefficients are obtained finally after some calculation 
\begin{equation}
\label{37}C_0=5K_c+2K_c(\epsilon _1-\epsilon _2-K_c\eta )[1+K_c\epsilon
_1/(k^2-4K_c)]^{-1}, 
\end{equation}
\begin{equation}
\label{38}C_1=-2aH\{K_c\epsilon _1+(1+\epsilon _1-\epsilon _2-K_c\eta
)[1+K_c\epsilon _1/(k^2-4K_c)]^{-1}\}. 
\end{equation}
The equation (\ref{36}) has the first order term $R^{\prime }$, we hope to
cancel it by introducing a new variable $\chi _k$, 
\begin{equation}
\label{39}\chi _k\equiv zR. 
\end{equation}
Thus equation (\ref{36}) becomes 
\begin{equation}
\label{42}\chi _k^{\prime \prime }+k^2\chi _k-C_0\chi _k-z^{-1}z^{\prime
\prime }\chi _k=0, 
\end{equation}
if we take 
\begin{equation}
\label{41}C_1=-2z^{-1}z^{\prime }. 
\end{equation}
Eq.(\ref{42}) is a standard equation without the damping term and with an
equivalent varying mass term $m_v^2=-C_0-z^{\prime \prime }/z$. Solving $z$
from (\ref{41}) and (\ref{38}), we get 
\begin{equation}
\label{43}z=-a\sqrt{2\epsilon _1[1+K_c\epsilon _1/(k^2-4K_c)]^{-1}}. 
\end{equation}
This is important parameter which controls evolution of the scalar
perturbation. Its first and second derivatives with respect to the conformal
time are given by 
\begin{equation}
\label{44}z^{\prime }/(aHz)=K_c\epsilon _1+(1+\epsilon _1-\epsilon
_2-K_c\eta )[1+K_c\epsilon _1/(k^2-4K_c)]^{-1}, 
\end{equation}
\begin{equation}
\label{45} 
\begin{array}{c}
z^{\prime \prime }/z=a^2H^2(k^2-4K_c+K_c\epsilon _1)^{-2}[k^4(2+2\epsilon
_1-3\epsilon _2+2\epsilon _1^2-4\epsilon _1\epsilon _2+\epsilon _3) \\ 
+K_ck^2(-16-12\epsilon _1+24\epsilon _2-15\epsilon _1^2+29\epsilon
_1\epsilon _2-8\epsilon _3-\epsilon _1^3+2\epsilon _1^2\epsilon _2-3\epsilon
_1\epsilon _2^2+\epsilon _1\epsilon _3) \\ 
+K_c^2(32+16\epsilon _1-48\epsilon _2+30\epsilon _1^2-52\epsilon _1\epsilon
_2+16\epsilon _3+3\epsilon _1^3-8\epsilon _1^2\epsilon _2+12\epsilon
_1\epsilon _2^2-4\epsilon _1\epsilon _3) \\ 
+K_c\eta (-68\epsilon _1-4k^4\epsilon _1+32\epsilon _2+2k^4\epsilon
_2-7\epsilon _1^2+16\epsilon _1\epsilon _2)+K_ck^2\eta ^2(-16-\epsilon _1)
\\ 
+K_c^2k^2\eta (33\epsilon _1-16\epsilon _2+2\epsilon _1^2-4\epsilon
_1\epsilon _2)+K_c^2\eta ^2(32+2k^4+4\epsilon _1)]. 
\end{array}
\end{equation}
The equation (\ref{36}) or (\ref{42}) becomes finally 
\begin{equation}
\label{46}\chi _k^{\prime \prime }+[k^2-3K_c-2K_cz^{\prime }/(aHz)-z^{\prime
\prime }/z]\chi _k=0, 
\end{equation}
with an equivalent varying mass $[-3K_c-2K_cz^{\prime }/(aHz)-z^{\prime
\prime }/z]^{1/2}$. Its spectrum, referring \cite{L-W} and \cite{G-L-T}, is 
\begin{equation}
\label{47}P_k^R=\frac{k(k^2-K_c)}{2\pi ^2}\frac{|\chi _k|^2}{z^2}. 
\end{equation}
If the universe is closed, the parameter $k$, almost to be dimensionless
comoving wavenumber, takes integer value $3$, $4$, $\cdots $, to infinity.
It is notable that the results gotten until now is exact, especially for (%
\ref{44}) and (\ref{45}), i.e., there is no any approximation to be taken.
Of course a precondition of such statement is to recognize the validity of
the first order approximation for the perturbative Einstein equation. If $%
K_c=0$, all return to standard result for flat case, i.e., (\ref{(1)}), (\ref
{(2)}) and (\ref{(3)}), specially $z\rightarrow z_0$. Eqs.(\ref{43})-(\ref
{46}) are our main results, which are the foundation for further analysis.
In fact, these results were able to be obtained almost by Refs.\cite{G-L-T}
and \cite{GMST}, if they would like to finish their last steps. However, we
adopt a rather different method to obtain the exact evolution equation of
the comoving curvature perturbation for non-flat case, this increases the
reliability of the results. The exact evolution equation of the tensor
perturbation for non-flat universe will be found in a forthcoming work.

\section{A try to treat new evolution equation}

{\hspace*{5mm}}The evolution equation obtained by us for non-flat case is
obviously more complicated than flat one. New parameter $\eta $ appears in
an unmanageable way. How to treat it brings us a new challenge. What is a
boundary condition on this equation? How to define the vacuum for this
quantized system? Noted that $\eta =a^{-2}H^{-2}\simeq \tau ^2$, the
coefficient of $\chi _k$ term, i.e., variable mass square, not only contains
a single term of $\tau ^{-2}$ like the flat case, but also contains other
terms which is very complicate function of the conformal time $\tau $ and
even various $\epsilon $ varied slowly. It is difficult to apply the
slow-roll method in this case. We must think out some better methods to
treat it. However, before finding out better outlet, we still hope to do a
comparison if we father upon so-called slow-roll approximation on it. In
such way we can appraise how the slow-roll approximation is partly invalid.

Let us introduce the potential parameters which are very suitable for
slow-roll approximation, seeing Ref.\cite{S-G}, 
\begin{equation}
\label{52}u_1\equiv (V_{,\phi }/V)^2,\quad \quad v_1\equiv V_{,\phi \phi
}/V,\quad \quad v_2\equiv V_{,\phi }V_{,\phi \phi \phi }/V^2, 
\end{equation}
to express the rolling parameters mentioned in (\ref{12}), 
\begin{equation}
\label{52-1} 
\begin{array}{c}
\epsilon _1\simeq \frac 12u_1-\frac 13u_1^2+\frac 13u_1v_1+ 
\frac{14}9K_cu_1\eta +\frac 49u_1^3-\frac 56u_1^2v_1+\frac 5{18}u_1v_1+\frac
19u_1v_2 \\ -\frac{67}{27}K_cu_1^2\eta +\frac{53}{27}K_cu_1v_1\eta +\frac{%
190 }{81}K_c^2u_1\eta ^2, 
\end{array}
\end{equation}
\begin{equation}
\label{52-2} 
\begin{array}{c}
\epsilon _2\simeq \frac 12u_1-v_1-\frac 53K_c\eta -\frac 23u_1^2+\frac
43u_1v_1-\frac 13v_1^2-\frac 13v_2+3K_cu_1\eta - 
\frac{20}9K_cv_1\eta -\frac{22}{27}K_c^2\eta ^2 \\ +\frac 32u_1^3-4u_1^2v_1+ 
\frac{23}9u_1v_1^2-\frac 29v_1^3+\frac{17}{18}u_1v_2-\frac 23v_1v_2-\frac
19v_3-\frac{167}{27}K_cu_1^2\eta \\ +\frac{70}9K_cu_1v_1\eta -K_cv_1^2\eta
-\frac 79K_cv_2\eta +\frac{166}{27}K_c^2u_1\eta ^2-\frac{49}{81}K_c^2v_1\eta
^2+\frac 8{81}K_c\eta ^3, 
\end{array}
\end{equation}
\begin{equation}
\label{52-3} 
\begin{array}{c}
\epsilon _3\simeq 2K_c\eta +u_1^2-\frac 52u_1v_1+v_1^2+v_2- 
\frac{13}3K_cu_1\eta +\frac{11}3K_cv_1\eta \\ - 
\frac{19}6u_1^3+\frac{55}6u_1^2v_1-\frac{13}2u_1v_1^2+\frac 23v_1^3-\frac
52u_1v_2+2v_1v_2+\frac 13v_3+\frac{67}6K_cu_1^2\eta -\frac{145}%
9K_cu_1v_1\eta \\ +\frac{29}9K_cv_1^2\eta +\frac 73K_cv_2\eta +\frac{22}%
9K_c^2\eta ^2-\frac{106}9K_c^2u_1\eta ^2+\frac{23}9K_c^2v_1\eta ^2+\frac{26}{%
81}K_c\eta ^3, 
\end{array}
\end{equation}
the terms with $K_c$ are our new results. Then we have, up to the second
order, 
\begin{equation}
\label{53} 
\begin{array}{c}
z^{\prime }/(aHz)\simeq 1+v_1+\dfrac 23K_c\eta -(\dfrac 12K_cu_1v_1+\dfrac
13K_c^2u_1\eta )/(k^2-4K_c) \\ 
+\dfrac 19(3u_1^2-9u_1v_1+3v_1^2+3v_2-13K_cu_1\eta +20K_cv_1\eta +\dfrac{22}%
3K_c^2\eta ^2), 
\end{array}
\end{equation}
\begin{equation}
\label{54} 
\begin{array}{c}
z^{\prime \prime }/z\simeq a^2H^2[(2+7K_c\eta -\dfrac 12u_1+3v_1)-(\dfrac
32K_cu_1v_1+2K_c^2u_1\eta )/(k^2-4K_c) \\ 
+\dfrac 1{18}(33u_1^2-69u_1v_1+36v_1^2+36v_2-142K_cu_1\eta +150K_cv_1\eta
+64K_c^2\eta ^2)]. 
\end{array}
\end{equation}
Moreover for the conformal time we have an approximation if the rolling
parameters are small 
\begin{equation}
\label{55}\tau \simeq -a^{-1}H^{-1}[1+(\epsilon _1-\frac 13K_c\eta
)+(3\epsilon _1^2-2\epsilon _1\epsilon _2-\frac 43K_c\epsilon _1\eta +\frac
15K_c^2\eta ^2)], 
\end{equation}
this result can be checked by $d\tau /dt=a^{-1}$ up to the appropriate
orders by using Eqs.(\ref{14}) and (\ref{15}). If we take the slow-roll
approximation, we have 
\begin{equation}
\label{56}\tau =-\frac{1+\mu }{aH}\simeq -\frac 1{aH}(1+\frac 12u_1-\frac
1{12}u_1^2+\frac 43u_1v_1-\frac 13K_c\eta +\frac{23}9K_cu_1\eta ), 
\end{equation}
then we get 
\begin{equation}
\label{57} 
\begin{array}{c}
z^{\prime \prime }/z\simeq \tau ^{-2}[(2+ 
\dfrac{17}3K_c\eta +\dfrac 32u_1+3v_1)-(\dfrac 32K_cu_1v_1+2K_c^2u_1\eta
)/(k^2-4K_c) \\ +(\dfrac 32u_1^2+\dfrac 92u_1v_1+2v_1^2+2v_2+9K_cu_1\eta + 
\dfrac{19}3K_cv_1\eta -\dfrac 4{45}K_c^2\eta ^2)]. 
\end{array}
\end{equation}
The equation (\ref{46}) is approximated as 
\begin{equation}
\label{58} 
\begin{array}{c}
\chi _k^{\prime \prime }+\{k^2- 
\dfrac{32}3K_c+K_c\cdot O(u_1,v_i)-\dfrac 1{\tau ^2}[2+\dfrac
32u_1+3v_1+\dfrac 32u_1^2+\dfrac 92u_1v_1+2v_1^2 \\ +2v_2-\dfrac{3K_cu_1v_1}{%
2(k^2-4K_c)}]+K_c\cdot (\tau ^2\text{ or higher terms})\}\chi _k=0. 
\end{array}
\end{equation}
Notice that we shall omit the $O(u_1,v_i)$ terms in the term of $\tau ^0$
order, and omit the higher terms such as $\tau ^2$, since they should be
small in the limit case $|$$\tau |\rightarrow 0$. Thus we obtained an
equation very similar with flat case. Since we omitted the $\tau ^2$ and
higher terms, it is obvious that we are not able to do analysis on the limit
case of $|\tau |\rightarrow \infty $. We then see that an Bessel asymptotic
analysis loses its validity.

If a Bessel analysis is forcedly adopted for the equation (\ref{58}), then
we have the spectrum 
\begin{equation}
\label{59}P_k^R=f(k)\dfrac{H^4}{4\pi ^2\dot \phi ^2}\cdot 2^{2\beta -3} 
\dfrac{\Gamma ^2(\beta )}{\Gamma ^2(3/2)}(1+\mu )^{1-2\beta }, 
\end{equation}
where 
\begin{equation}
\label{60}f(k)=k(k^2-K_c)(k^2-\dfrac{32}3K_c)^{-3/2}[1+K_c\epsilon
_1/(k^2-4K_c)], 
\end{equation}
and 
\begin{equation}
\label{61}\beta =\dfrac 32+\frac 12u_1+v_1+\frac 5{12}u_1^2+\frac
76u_1v_1+\frac 13v_1^2+\frac 23v_2-\dfrac 12K_cu_1v_1/(k^2-4K_c). 
\end{equation}
It is unnecessary for us to go further, not due to mathematical difficulty,
but since the evolution is not slow-roll in the small $k$ region, where is
of greatest interest for us. In other words, we don't reassure our
approximation, in spite of it is very efficient for flat case. However,
these equations still give us some elicitation. We have seen that there will
be some change in the forthcoming formulae of the spectral index and its
running, which will be different from the flat case. Even if the universe is
slight curved, the slow-rolling is still a good approximation for the power
spectrum of the small angle part.

We must note the relationship between the dimensionless comoving wavenumber $%
k$ and the multipole $l$ in CMBR spectrum if the universe is closed. If the
horizon of our universe is about the present Hubble distance $H_0^{-1}$, the
length corresponded to the multipole $l$ is about $\pi H_0^{-1}/l$,
therefore this length should be equal to the physical half-wavelength $\pi $$%
a_0/k$ corresponded to wavenumber $k$, thus we have $k\sim a_0H_0l$, where $%
a_0$ is the present cosmic scale factor with length dimensional. If the
total density $\Omega _T$ is about $1.01$, then $a_0H_0\simeq 10$. For the
quadrupole fluctuation of $l=2$ in CMBR, which has the real cosmological
signification, the wavenumber $k$ is about $20$. Therefore the enhanced
effect of the factor $f(k)$ for spectrum is not too large. The leading
behavior of the spectrum is controlled by the parameter $z\sim z_0$, 
\begin{equation}
\label{62}k^3\chi ^2\sim \eta ^{-1},\quad \quad P_k^R\sim z^{-2}\eta
^{-1}\sim H^4/\dot \phi _0^2, 
\end{equation}
i.e., the factor $H^4/\dot \phi _0^2$ in Eq.(\ref{59}) governs the spectrum,
just the same as the flat case. This key feature has been captured by Ref.%
\cite{L-D}, specially seeing their figure 15. They think this can explain
the low power in small multipole $l$ for CMBR spectrum (figure 20 in \cite
{L-D}).

If the universe is born from nothing to a finite size one through a
tunnelling effect, as described by a model of Ref.\cite{Zhang}, at this
beginning the universe has an equilibrium of $K_ca^{-2}=V/3$ from Eq.(\ref{4}%
), we then have a asymptotic behavior near the time starting point $%
t\rightarrow 0$ 
\begin{equation}
\label{63}H\sim \dot \phi \sim t,\quad \quad H^4/\dot \phi ^2\sim t^2. 
\end{equation}
The quantity $H^4/\dot \phi ^2$ will increase from zero up to its maximum at
some time point, keeping this high value for a period with a very slowly
decreasing, i.e., an approximated de Sitter phase, then fall off to end its
inflation. Before this time point the spectrum is suddenly low for small $k$%
, i.e., for a sky large angle or low multipole $l$, looks like a cut-off. It
seems possible that the closed universe can explain the large angle lack of
the CMBR spectrum and the little up-departure of the cosmic total density
from the exact flat case. Anyhow we must remember that the result (\ref{58}%
)-(\ref{61}) obtained by us is only a try. In order to obtain an assured
consequence, it is necessary to investigate numerically on the exact
evolution equation (\ref{43})-(\ref{47}), which will be our further serious
tasks.

\section{Conclusion}

{\hspace*{5mm}}We obtained an exact evolution equation of the scalar
perturbation for non-flat universe by using a direct method. The exact
evolution equation of the scalar perturbation for flat universe has been
playing an important role in the getting a primordial inflation spectrum.
Its non-flat extension will play a key role in future to analysis whether
the universe is closed or flat. The idea of the closed universe is
attractive. After all, this idea is the simplest one among the many ideas to
introduce a new scale. We should start with our investigation from as simple
idea as possible.

The exact evolution equation obtained by us is rather complicated. We have
to do a try, in spite of which is immature, however, surely beneficial. It
is impossible that the slow-roll approximation will lost wholly its
efficiency even in non-flat case. Maybe some revelation can be drawn out
from this rather simplified treatment.

\bigskip

\noindent {\bf Acknowledgments}\\ The project supported by National Natural
Science Foundation of China under Grant Nos 10047004 and NKBRSP G19990754.


\end{document}